\documentstyle[aps,twocolumn,floats,epsfig]{revtex}
\begin{document}
\title{Josephson effect in point contacts between ''f-wave'' superconductors. }
\author{R. Mahmoodi $^{(1)}$, S.N. Shevchenko$^{(2)}$, Yu.A. Kolesnichenko$^{(1,2)}$}
\address{$^{\left( 1\right) }$ ${}$ Institute for Advanced Studies in Basic Sciences,%
\\
45195-159, Gava Zang, Zanjan, Iran \\
$^{\left( 2\right) }$\ B.Verkin Institute for Low Temperature Physics and\\
Engineering, \\
National Academy of Sciences of Ukraine, \\
47 Lenin Ave., 61103 Kharkov, Ukraine}
\date{\today}
\maketitle

\begin{abstract}
A stationary Josephson effect in point contacts between triplet
superconductors is analyzed theoretically for  most probable
models of the order parameter in $UPt_{3}$ and $Sr_{2}RuO_{4}.$
The consequence of misorientation of crystals in superconducting
banks on this effect is considered. We show that different models
for the order parameter lead to quite different current-phase
dependences. For certain angles of misorientation a boundary
between superconductors can generate the parallel to surface
spontaneous current. In a number of cases the state with a zero
Josephson current and minimum of the free energy corresponds to a
spontaneous phase difference. This phase difference depends on
the misorientation angle and may possess any value. We conclude
that experimental investigations of the current-phase dependences
of small junctions can be used for determination of the order
parameter symmetry in the mentioned above superconductors.
\end{abstract}

\section{\protect\bigskip Introduction.}

Triplet superconductivity, which is an analogue of superfluidity in $%
^{3}He, $ was firstly discovered in a heavy-fermion compound
$UPt_{3}$ more than ten years ago \cite{Muller,Qian}. Recently,
a novel triplet superconductor $%
Sr_{2}RuO_{4}$ was found \cite{Maeno1,Maeno2}. In these
compounds, the triplet pairing can be reliably determined, for
example, by Knight shift experiments \cite{Tou,Ishida}, however
the identification of a symmetry of the order parameter is much
more difficult task. A large number of experimental and
theoretical investigations done on $UPt_{3}$ and $Sr_{2}RuO_{4}$
are concerned with different thermodynamic and transport
properties, but the precise order parameter symmetry remains
still to be worked out yet (see, for example, \cite
{Mashida,Graf,Dahm,Graf1}, and original references therein).

Calculations of the order parameter $\widehat{\Delta }\left( \widehat{%
{\bf k}}\right) $ in$\ UPt_{3}$ and $Sr_{2}RuO_{4}$ as a function
of the momentum direction  $\widehat{{\bf k}}$ on the Fermi
surface is a very complex problem. Some general information about
$\widehat{\Delta }\left( \widehat{{\bf k}}\right)$ can be
obtained from a symmetry of a normal state: $G_{spin-orbit} \times
{\huge \tau} \times\ U(1)$, where $G_{spin-orbit}$ represents the
point group with inversion; ${\huge \tau}$ is the time-inversion
operator, and $U(1)$ is the group of gauge transformation. A
superconducting state breaks one or more symmetries. In
particular, a transition to the superconducting state implies an
appearance of the phase coherence corresponding to breaking of
the gauge symmetry. According to  Landau theory \cite{Landau} of
second order phase transitions, the order parameter is
transformed only on irreducible representations of the symmetry
group of the normal state. Conventional superconducting states
have a total point symmetry of the crystal and belong to the
unitary even representation $A_{1g}.$ In unconventional
superconductors this symmetry is broken. The parity of a
superconductor with inversion symmetry can be specified using the
Pauli principle. Because for triplet pairing the spin part of the
$\widehat{\Delta }$ is a symmetric second rank spinor, the
orbital part has to belong to the odd representation. In a
general case the triplet paring is described by the order
parameter of the form $\widehat{\mathbf{\Delta }}\left(
\widehat{\mathbf{k}}\right) =i\mathbf{d}%
\left( \widehat{\mathbf{k}}\right) \widehat{\mathbf{\sigma
}}\widehat{\sigma }_{2}$, where the vector $\widehat{\sigma }%
=\left( \widehat{\sigma }_{1},\widehat{\sigma }_{2},\widehat{\sigma }%
_{3}\right) ,$ and $\widehat{\sigma }_{i}$ are Pauli matrices
in the spin space. A vector $\mathbf{d}%
\left( \widehat{\mathbf{k}}\right)=-\mathbf{d}%
\left( -\widehat{\mathbf{k}}\right)$ in spin space is frequently
referred to as an order parameter or a gap vector of the triplet
superconductor. This vector defines the axis along which the
Cooper pairs have zero spin projection. If $\mathbf{d}$ is
complex, the spin components of the order parameter spontaneously
break time-reversal symmetry.

Symmetry considerations reserves for the order parameter
considerable freedom in the selection of irreducible
representation and its basis functions. Therefore in many papers
(see, for example, \cite
{Mashida,Graf,Dahm,Graf1,Graf2,Sigrist,Won}) authors consider
different
models (so-called scenarios) of superconductivity in $UPt_{3}$ and $%
Sr_{2}RuO_{4},$ which are based on possible representations of
crystallographic point groups.
The subsequent comparison of
theoretical results with experimental data makes it possible to conclude
on the symmetry of the order parameter.

In real crystalline superconductors there is no classification of
Cooper pairing by angular momentum ($s$-wave, $p$-wave, $d$-wave,
$f$-wave pairing, etc.). However these terms are often used for
unconventional superconductors meaning that the point symmetry of
the order parameter is the same as one for the corresponding
representation of $SO_{3}$ symmetry group of isotropic conductor.
In this terminology conventional superconductors can be referred
to as $s$-wave. For example, the ''$p$-wave'' pairing corresponds
to the odd two-dimensional representation $E_{1u}$ of $D_{6h}$ point group or $%
E_{u}$ representation of $D_{4h}$ point group. The order parameter
for these representations has the same symmetry, as for the
superconducting state with angular momentum $l=1$ of Cooper pairs
in an isotropic conductor. If the symmetry of $\widehat{\Delta }$
can not be formally related to any irreducible representation of
$SO_{3}$ group, these states are usually referred to as hybrid
states.

Apparently, in crystalline triplet superconductors the order
parameter has more complex dependence on $\widehat{{\bf k}}$ in
comparison with well known $p$-wave order parameter for superfluid
phases of $^{3}He.$ The heavy-fermion superconductor $UPt_{3}$
belongs to the hexagonal crystallographic point group $\left(
D_{6h}\right) ,$ and it is most likely that the pairing state
belongs to $E_{2u}$ (''$f-$wave'' state) representation. A
layered perovskite material $Sr_{2}RuO_{4}$ belongs to the
tetragonal crystallographic point group $\left( D_{4h}\right)$ .
Initially the simplest ''$p$-wave'' model based on $E_{u}$
representation was proposed for the superconducting state in this
compound \cite{Rice,Agt}. However this model was inconsistent with
available experimental data, and later \cite {Graf,Dahm} other
''$f$-wave'' models of pairing state were proposed.

Theoretical studies of specific heat, thermal conductivity,
ultrasound absorption for different models of triplet
superconductivity show considerable quantitative differences
between calculated dependences\cite {Mashida,Graf,Dahm,Won}. The
Josephson effect is much more sensitive to  dependence of
$\widehat{\Delta }$ on the momentum direction on the Fermi
surface. One of the possibilities to form a Josephson junction is
to create a point contact between two massive superconductors. A
microscopic theory of the stationary Josephson effect in
ballistic point contacts between conventional superconductors was
developed in Ref.\cite{KO}. Later this theory was generalized for
a pinhole model in $^{3}He$ \cite{Kurk,Yip} and for point contacts
between ''$d$-wave'' high-$T_{c}$ superconductors
\cite{AmOmZ,Fogelstrom}. It was shown that current-phase
dependences for the Josephson current in such systems are quite
different from those of conventional superconductors, and states
with a spontaneous phase difference become possible. Theoretical
and experimental investigations of this effect in novel triplet
superconductors seem to be interesting and enable one to
distinguish among different candidates for the superconducting
state.

In Ref.\cite{BBF01} the authors study the interface Andreev bound
states and their influence on the Josephson current between clean
''$f$-wave'' superconductors both self-consistently (numerically)
and non-self-consistently (analytically). The temperature
dependence of the critical current is presented. However in that
paper there is no detailed analysis of the current-phase
dependences for different orientations of the crystals in the
superconducting banks.

In this paper we theoretically investigate the stationary
Josephson effect in a small ballistic junction between two bulk
triplet superconductors with different  orientations of the
crystallographic axes with respect to the junction normal. In
Sec.II we describe our model of the junction and present the full
set of equations. In Sec.III the current density in the junction
plane is analytically calculated for a non-self-consistent model
of the order parameter. In Sec.IV the current-phase dependences
for most likely models of ''$f$-wave'' superconductivity in
$UPt_{3}$ and $Sr_{2}RuO_{4}$ are analyzed for different mutual
orientations of the banks. We end in Sec.V with some conclusions.

\section{Model of the contact and formulation of the problem.}

We consider a model of a ballistic point contact as an orifice
with a diameter $d$ in an impenetrable for electrons partition
between two superconducting half spaces (Fig.1). We assume that
the contact diameter $d$ is much larger than the Fermi wavelength
and use the quasiclassical approach.
\begin{figure}[t]
\epsfysize 6cm \epsfbox[0 250 500 620]{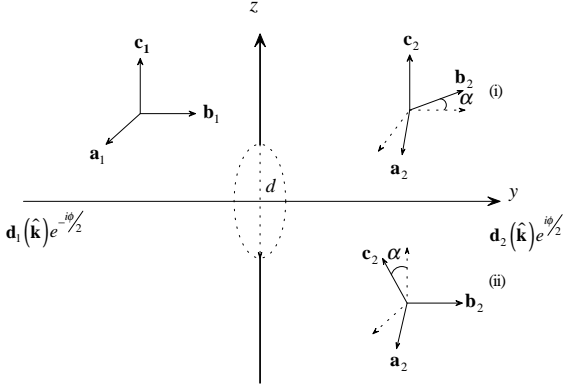}\caption{Scheme of
a contact in the form of an orifice between two superconducting
banks, which are misorientated on an angle $\protect\alpha $.}
\label{Fig1}
\end{figure}
In order to calculate the stationary Josephson current in point
contact we use ''transport-like'' equations for $\xi $-integrated
Green functions $\stackrel{\vee }{g}\left( \widehat{{\bf k}},{\bf r},\varepsilon _{m}\right)
$\cite{Eilen}%
\begin{equation}
\left[ i\varepsilon _{m}\stackrel{\vee }{\tau }_{3}-\stackrel{\vee }{\Delta }%
,\stackrel{\vee }{g}\right] +iv_{F}\widehat{{\bf k}}\nabla \stackrel{\vee }{g%
}=0,  \label{quasiclass_Eq}
\end{equation}
and the normalization condition
\begin{equation}
\stackrel{\vee }{g}\stackrel{\vee }{g}=-1.  \label{norm_cond}
\end{equation}
Here $\varepsilon _{m}=\pi T(2m+1)$ are discrete Matsubara
energies, $v_{F}$ is the Fermi velocity, $\widehat{{\bf k}}$
is a unit vector along the electron velocity, $\stackrel{\vee }{\tau }_{3}=%
\widehat{\tau }_{3}\otimes \widehat{I}; \widehat{\tau }_{i} \left(
i=1,2,3\right)$ are Pauli matrices in a particle-hole space.

The Matsubara propagator $\stackrel{\vee }{g}$ can be written in
the form \cite{Ser}:
\begin{equation}
\stackrel{\vee }{g}=\left(
\begin{array}{cc}
g_{1}+{\bf g}_{1}\widehat{\sigma } & \left( g_{2}+{\bf g}_{2}\widehat{{\bf %
\sigma }}\right) i\widehat{\sigma }_{2} \\
i\widehat{\sigma }_{2}\left( g_{3}+{\bf g}_{3}\widehat{{\bf \sigma }}\right)
& g_{4}-\widehat{\sigma }_{2}{\bf g}_{4}\widehat{{\bf \sigma }}\widehat{{\bf %
\sigma }}_{2}
\end{array}
\right) ;
\end{equation}
as can be done for an arbitrary Nambu matrix. Matrix structure of the off-diagonal
self energy $%
\stackrel{\vee }{\Delta }$ in Nambu space is
\begin{equation}
\stackrel{\vee }{\Delta }=\left(
\begin{array}{cc}
0 & i{\bf d}\widehat{{\bf \sigma }}\widehat{{\bf \sigma }}_{2} \\
i\widehat{\sigma }_{2}{\bf d}^{\ast }\widehat{\sigma } & 0
\end{array}
\right) ,
\end{equation}
Below we consider so-called unitary states, for which ${\bf d\times d}%
^{\ast }=0.$ \

The gap vector ${\bf d}$ has to be determined from the
self-consistency equation:
\begin{equation}
{\bf d}\left( \widehat{{\bf k}},{\bf r}\right) =\pi TN\left( 0\right)
\sum_{m}\left\langle V\left( \widehat{{\bf k}},\widehat{{\bf k}}^{\prime
}\right) {\bf g}_{2}\left( \widehat{{\bf k}}^{\prime },{\bf r},\varepsilon
_{m}\right) \right\rangle ,  \label{self-cons_Eq}
\end{equation}
where $V\left( \widehat{{\bf k}},\widehat{{\bf k}}^{\prime }\right) $ is a
potential of pairing interaction; $\left\langle ...\right\rangle $ stands
for averaging over directions of an electron momentum on the Fermi surface; $%
N\left( 0\right) $ is the electron density of states.

Solutions of Eqs. (\ref{quasiclass_Eq}), (\ref{self-cons_Eq})
must satisfy the conditions for Green functions and vector $\bf d$
in the banks of superconductors far from the orifice:
\begin{eqnarray}
\stackrel{\vee }{g}\left( \mp \infty \right) &=&\frac{i\varepsilon _{m}%
\stackrel{\vee }{\tau }_{3}-\stackrel{\vee }{\Delta }_{1,2}}{\sqrt{%
\varepsilon _{m}^{2}+\left| {\bf d}_{1,2}\right| ^{2}}};  \label{g(inf)} \\
{\bf d}\left( \mp \infty \right) &=&{\bf d}_{1,2}\left( \widehat{{\bf k}}%
\right) \exp \left( \mp \frac{i\phi }{2}\right) ,  \label{d(inf)}
\end{eqnarray}
where $\phi $ is the external phase difference. Eqs. (\ref{quasiclass_Eq}) and (%
\ref{self-cons_Eq}) have to  be supplemented by the boundary
continuity conditions at the contact plane and conditions of
reflection at the interface between superconductors. Below we
assume that this interface is smooth and  electron scattering is
negligible.

\section{Calculation of the current density.}

The solution of Eqs. (\ref{quasiclass_Eq}) and (\ref{self-cons_Eq}) allows
us to calculate the current density:
\begin{equation}
{\bf j}\left( {\bf r}\right) =2\pi eTv_{F}N\left( 0\right)
\sum_{m}\left\langle \widehat{{\bf k}}g_{1}\left( \widehat{{\bf k}},{\bf r}%
,\varepsilon _{m}\right) \right\rangle .  \label{j(r)}
\end{equation}

We  consider a simple model of the constant order parameter
up to the surface. The pair breaking and the scattering on the
partition and in the junction are ignored. This model can be
rigorously found for the calculations of the current density
(\ref{j(r)}) in ballistic point contacts
between conventional superconductors at zero approximation with
  small parameter $%
d/\xi _{0}$ ($\xi _{0}$ is the coherence length) \cite{KO}. In
anisotropically paired superconductors the order parameter
changes at  distances of the order of $\xi _{0}$ even near a
specular surface \cite {Buch,Mats}. Thus for calculations of the
current (\ref{j(r)}) in the main approximation on the parameter
$d/\xi _{0}$ it is necessary to solve the equation
(\ref{self-cons_Eq}) near a surface of the semi-infinite
superconductor. It can be done only numerically and will be the
subject of our future investigations. Below we assume that the
order parameter does not depend on coordinates and in each
half-space equals to its value (\ref{d(inf)}) far form the point
contact. For this non-self-consistent model the current-phase
dependence of a Josephson junction can be calculated analytically.
It makes possible to analyze the main features of current-phase
dependences for different scenarios of ''$f$-wave''
superconductivity. We believe that under this strong assumption
our results describe the real situation qualitatively, as it was
justified for point contacts between ''$d$-wave'' superconductors
\cite{AmOmZ} and pinholes in $^{3}He$ \cite{Viljas}. It was also
shown in Ref.\cite{BBF01} that for the contact between
''$f$-wave'' superconductors there is also good qualitative
agreement between self-consistent and non-self-consistent
solutions (although, of course, quantitative distinctions, take
place).

In a ballistic case the system of 16 equations for functions $g_{i}$ and $%
{\bf g}_{i}$ can be decomposed on independent blocks of
equations. The set of equations which enables us to find the Green
function $g_{1}$ is
\begin{eqnarray}
iv_{F}\widehat{{\bf k}}\nabla g_{1}+\left( {\bf g}_{3}{\bf d}-{\bf g}_{2}%
{\bf d}^{\ast }\right) &=&0;  \label{a} \\
iv_{F}\widehat{{\bf k}}\nabla {\bf g}_{-}+2i\left( {\bf d\times g}_{3}+{\bf d%
}^{\ast }{\bf \times g}_{2}\right) &=&0;  \label{b} \\
iv_{F}\widehat{{\bf k}}\nabla {\bf g}_{3}-2i\varepsilon _{m}{\bf g}%
_{3}-2g_{1}{\bf d}^{\ast }-i{\bf d}^{\ast }\times {\bf g}_{-} &=&0;
\label{c} \\
iv_{F}\widehat{{\bf k}}\nabla {\bf g}_{2}+2i\varepsilon _{m}{\bf g}%
_{2}+2g_{1}{\bf d}-i{\bf d}\times {\bf g}_{-} &=&0;  \label{d}
\end{eqnarray}
where ${\bf g}_{-}={\bf g}_{1}-{\bf g}_{4}.$ The Eqs.
(\ref{a})-(\ref{d}) can be solved by integrating over ballistic
trajectories  of electrons in the\ right and left half-spaces.
The general solution satisfying the boundary conditions
(\ref{g(inf)}) at infinity is
\begin{eqnarray}
g_{1}^{\left( n\right) } &=&\frac{i\varepsilon _{m}}{\Omega _{n}}+iC_{n}\exp
\left( -2s\Omega _{n}t\right) ;  \label{e} \\
{\bf g}_{-}^{\left( n\right) } &=&{\bf C}_{n}\exp \left( -2s\Omega
_{n}t\right) ;  \label{f} \\
{\bf g}_{2}^{\left( n\right) } &=&-\frac{2C_{n}{\bf d}_{n}-{\bf d}_{n}\times
{\bf C}_{n}}{-2s\eta \Omega _{n}+2\varepsilon _{m}}\exp \left( -2s\Omega
_{n}t\right) -\frac{{\bf d}_{n}}{\Omega _{n}};  \label{g} \\
{\bf g}_{3}^{\left( n\right) } &=&\frac{2C_{n}{\bf d}_{n}^{\ast }+{\bf d}%
_{n}^{\ast }\times {\bf C}_{n}}{-2s\eta \Omega _{n}-2\varepsilon _{m}}\exp
\left( -2s\Omega _{n}t\right) -\frac{{\bf d}_{n}^{\ast }}{\Omega _{n}};
\label{h}
\end{eqnarray}
where $t$ is the time of the flight along the trajectory, $sgn\left(
t\right) =sgn\left( z\right) =s;$ $\eta =sgn\left( v_{z}\right) ;$ $\Omega
_{n}=\sqrt{\varepsilon _{m}^{2}+\left| {\bf d}_{n}\right| ^{2}}.$ By
matching the solutions (\ref{e}-\ref{h}) at the orifice plane $\left(
t=0\right) $, we find constants $C_{n}$ and ${\bf C_{n}}.$ Index $n$ numbers left $%
\left( n=1\right) $ and right $\left( n=2\right) $ half-spaces.
The function $g_{1}\left( 0\right) =g_{1}^{\left( 1\right) }\left(
-0\right) =g_{1}^{\left( 2\right) }\left( +0\right) ,$ which
determines the current density in the contact is

\[g_{1}\left( 0\right) =\]
\begin{equation}
\frac{i\varepsilon _{m}\cos \varsigma \left( \Omega _{1}+\Omega
_{2}\right) +\eta \sin \varsigma \left( \varepsilon
_{m}^{2}+\Omega _{1}\Omega _{2}\right) }{{\bf \Delta }_{1}{\bf \Delta }%
_{2}+\left( \varepsilon _{m}^{2}+\Omega _{1}\Omega _{2}\right)
\cos \varsigma -i\varepsilon _{m}\eta \left( \Omega _{1}+\Omega
_{2}\right) \sin \varsigma }  \label{g1(0)}
\end{equation}
In the formula (\ref{g1(0)}) we have taken into account that for unitary
states vectors ${\bf d}_{1,2}$ can be written as
\begin{equation}
{\bf d}_{n}={\bf \Delta }_{n}\exp i\psi _{n},
\end{equation}
where ${\bf \Delta }_{1,2}$ are real vectors.

Knowing the function $g_{1}\left( 0\right)$ one can calculate the
current density at the orifice plane ${\bf j}(0)$:
\begin{equation}
{\bf j}(0)=4\pi eN(0)v_{F}T\sum_{m=0}^{\infty }\int d\widehat{{\bf k}}%
\widehat{{\bf k}}%
\mathop{\rm Re}%
g_{1}(0),  \label{jJos}
\end{equation}
where

Re $g_{1}\left( 0\right)=$

\begin{equation}
\frac{\sin \varsigma \left( \Delta _{1}^{2}\Delta _{2}^{2}\cos
\varsigma +\left( \varepsilon _{m}^{2}+\Omega _{1}\Omega
_{2}\right) {\bf \Delta }_{1}{\bf \Delta }_{2}\right) }{\left[ {\bf \Delta }%
_{1}{\bf \Delta }_{2}+\left( \varepsilon _{m}^{2}+\Omega
_{1}\Omega _{2}\right) \cos \varsigma \right] ^{2}+\varepsilon
_{m}^{2}\left( \Omega _{1}+\Omega _{2}\right) ^{2}\sin
^{2}\varsigma}
\end{equation}
or, alternatively,
\begin{equation}
\mathop{\rm Re}%
g_{1}\left( 0\right)=\frac{\Delta _{1}\Delta
_{2}}{2}\sum\limits_{\pm }\frac{\sin (\varsigma \pm \theta
)}{\varepsilon _{m}^{2}+\Omega _{1}\Omega _{2}+\Delta _{1}\Delta
_{2}\cos (\varsigma \pm \theta )}, \label{Re(g1)_2}
\end{equation}
where $\theta $ is defined by ${\bf \Delta }_{1}(\widehat{{\bf
k}}){\bf \Delta }_{2}(\widehat{{\bf k}})=\Delta
_{1}(\widehat{{\bf k}})\Delta _{2}(\widehat{{\bf k}})\cos \theta$, and $\varsigma (\widehat{{\bf k}})=\psi _{2}(%
\widehat{{\bf k}})-\psi _{1}(\widehat{{\bf k}})+\phi .$

Misorientation of the crystals would generally result in
appearance of the current along the interface\cite{AmOmZ,BBF01}
as can be calculated by projecting vector ${\bf j}$ at the
corresponding direction.

We consider a rotation $R$ only in the right superconductor (see,
Fig.1), (i.e., ${\bf d}_{2}\left( \widehat{{\bf k}}\right) =R{\bf
d}_{1}\left( R^{-1}\widehat{{\bf k}}\right)$). The $c$-axis in the
left half-space we choose along the partition between
superconductors (along $z$-axis in Fig.1). To illustrate results
obtained by computing the formula (\ref{jJos}), we plot the
current-phase dependence for different below mentioned scenarios
of '' $f$-wave'' superconductivity for two different geometries
corresponding to different orientations of the crystals to the
right and to the left at the interface (see, Fig.1):

(i) The basal $ab$-plane to the right is rotated about $c$-axis
by the angle $\alpha $;
$\widehat{{\bf c}}_{1}\Vert \widehat{{\bf c}}_{2}$.%

(ii) The $c$-axis to the right is rotated about the contact axis
($y$-axis in Fig.1) by the angle $\alpha$; $\widehat{{\bf
b}}_{1}\Vert \widehat{{\bf b}}_{2}$.

Further calculations require a certain model of the
vector order parameter ${\bf d}$.

\section{Current-phase dependence for different scenarios of ''$f$-wave''
superconductivity.}

The model which has been successful to explain
properties of the superconducting phases in $UPt_{3}$ is based on
the odd-parity $E_{2u}$ representation of the hexagonal point
group $D_{6h}$\ for the strong spin-orbital coupling with vector
{\bf d} locked along {\bf c} axis of the lattice \cite{Graf}:
${\bf d=}\Delta _{0}\widehat{z}\left[ \eta _{1}Y_{1}+\eta
_{2}Y_{2}\right] ,$ where $Y_{1}=k_{z}\left(
k_{x}^{2}-k_{y}^{2}\right) $ and $Y_{2}=$ $2k_{x}k_{y}k_{z}$ are
the basis
function of the representation\footnote{%
Strictly speaking, in crystals with a strong spin-orbit coupling
the spin is a ''bad'' quantum number, but electron states are
twice degenerated and can be characterized by pseudospins.}. The
coordinate axes $x, y, z$ here and below are chosen along the
crystallographic axes $\widehat{{\bf a}}, \widehat{{\bf b}},
\widehat{{\bf c}}$ as at the left at Fig.1. This model describes
the hexagonal analog of spin-triplet $f$-wave pairing. For the
high-temperature A-phase ($\eta _{2}=0$) the order parameter has
an equatorial line node and two longitudinal \ line nodes. In the
low-temperature B-phase ($\eta _{2}=i$) or the axial state:
\begin{equation}
{\bf d=}\Delta _{0}\widehat{z}k_{z}\left( k_{x}+ik_{y}\right) ^{2},
\label{axial}
\end{equation}
the longitudinal \ line nodes are closed and there is a pair of point nodes.
The B-phase (\ref{axial}) breaks the time-reversal symmetry. The function $\Delta _{0}=$ $%
\Delta _{0}\left( T\right) $ in Eq.\ref{axial} and below
describes the dependence of the order parameter ${\bf d}$ on
temperature $T$ (carrying out numerical calculations we
assume $T=0$).

\begin{figure}[t]
\epsfysize 6cm \epsfbox[50 50 100 550]{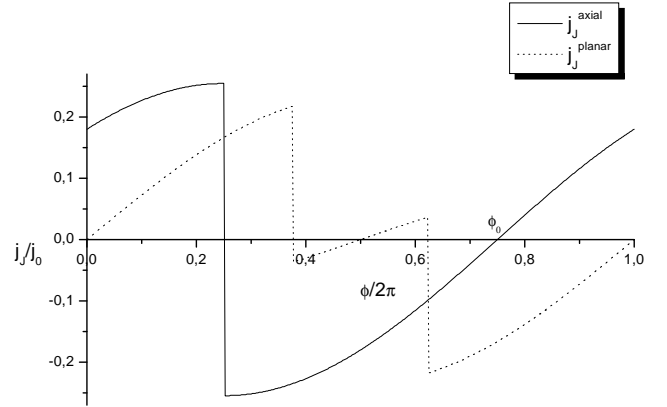}\caption{Josephson
current densities versus phase $\protect\phi $ for axial (22)
and planar (23) states in the geometry (i); misorientation angle
$\protect\alpha =\protect\pi /4$; the current is given in units of
$j_{0}=\frac{\protect\pi }{2} eN(0)v_{F}\Delta _{0}(0)$.}
\label{Fig2}
\end{figure}

Other candidates to describe the orbital states, which imply that
the effective spin-orbital coupling in $UPt_{3}$ is weak, are
unitary planar state
\begin{equation}
{\bf d=}\Delta _{0}k_{z}(\widehat{x}\left( k_{x}^{2}-k_{y}^{2}\right) +%
\widehat{y}2k_{x}k_{y}),  \label{planar}
\end{equation}
(or ${\bf d=}\Delta _{0}\left( Y_{1},Y_{2},0\right) $) and non-unitary
bipolar state ${\bf d=}\Delta _{0}\left( Y_{1},iY_{2},0\right) $ \cite
{Mashida}. In Fig.\ref{Fig2} we plot the Josephson current-phase dependence $%
j_{J}(\phi )=j_{y}(y=0)$ calculated from Eq.\ref{jJos} for both
the axial (with the order parameter given by Eq.\ref
{axial}) and the planar (Eq.\ref{planar}) states for a particular value of $%
\alpha $ under the rotation of basal $ab$-plane to the right (the
geometry (i)). For simplicity we use the spherical model of the
Fermi surface. For the axial state the current-phase dependence
is just the slanted sinusoid and for the planar state it shows a
''$\pi $-state''. The appearance of $\pi $-state\ at low
temperatures is due to the fact that different quasiparticle
trajectories contribute to the current with different effective phase differences $%
\varsigma (\widehat{{\bf k}})$\ (see Eqs.\ref{jJos} and
\ref{Re(g1)_2})\cite {Yip}. Such a different behavior can be a
criterion to distinguish between the axial and the planar states,
taking advantage of the phase-sensitive Josephson effect. Note
that for the axial model the Josephson current formally does not
equal to zero at $\phi =0.$ This state is unstable (does not
correspond to a minimum of the Josephson energy), and the state
with a spontaneous phase difference (value $\protect\phi_{0}$ in
Fig.2), which depends on the misorientation angle $\alpha ,$ is
realized.

\begin{figure}[t]
\epsfysize 5cm \epsfbox[60 50 100 500]{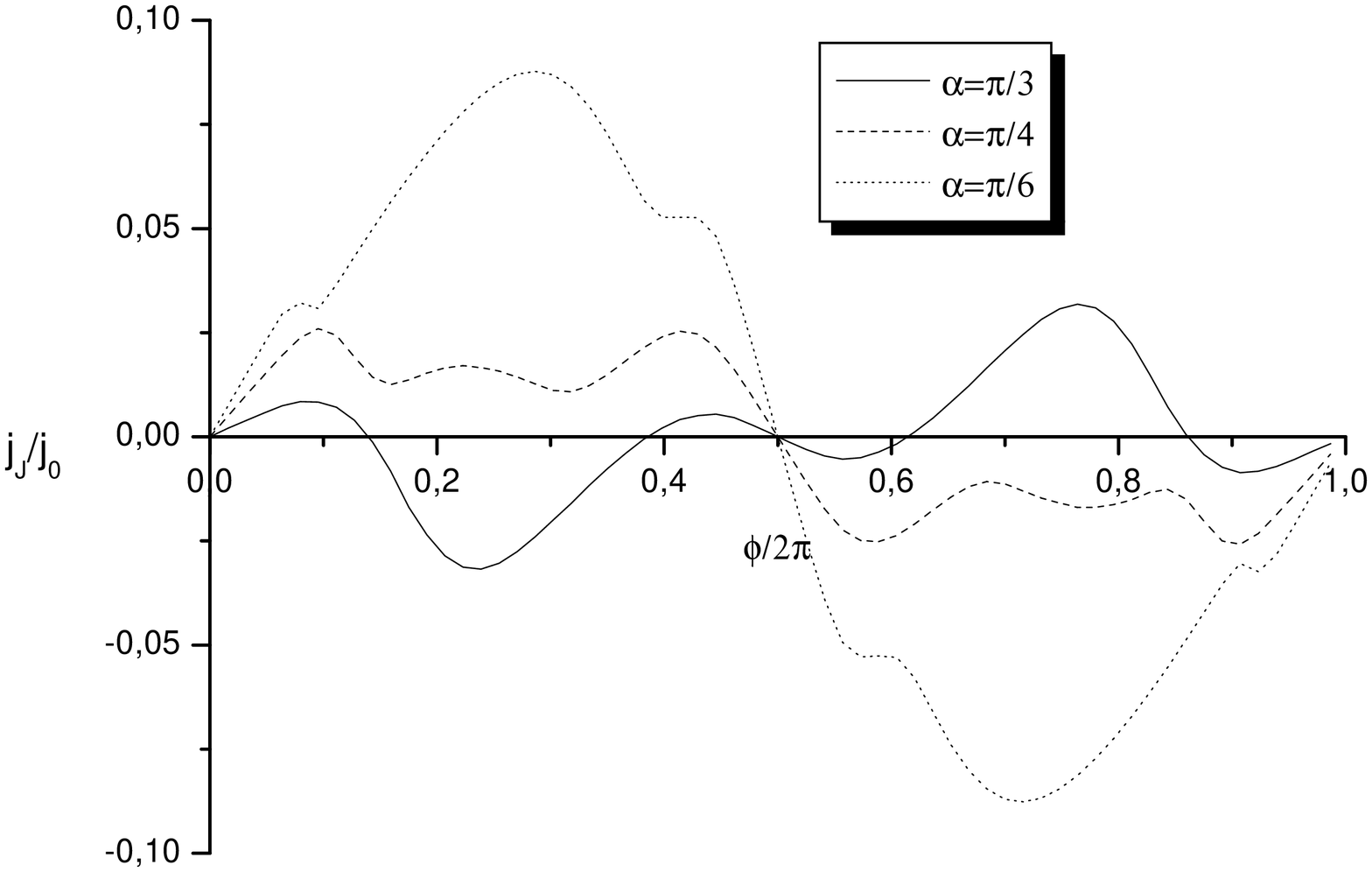}\caption{Josephson
current versus phase $\protect\phi $ for the axial (22) state in
the geometry (ii) for different $\protect\alpha $.} \label{Fig3}
\end{figure}

The remarkable influence of  the misorientation angle $\alpha $\
on the current-phase dependence is shown in Fig.\ref{Fig3} for the
axial state in the geometry (ii). For some values of $\alpha $ (in
Fig.\ref{Fig3}  it is $\alpha =\pi /3$ ) there are more than one
state, which correspond to  minima of the Josephson energy
$(j_J=0$ and $dj_J/d\phi>0)$.

\begin{figure}[t]
\epsfysize 6cm \epsfbox[60 0 50 550]{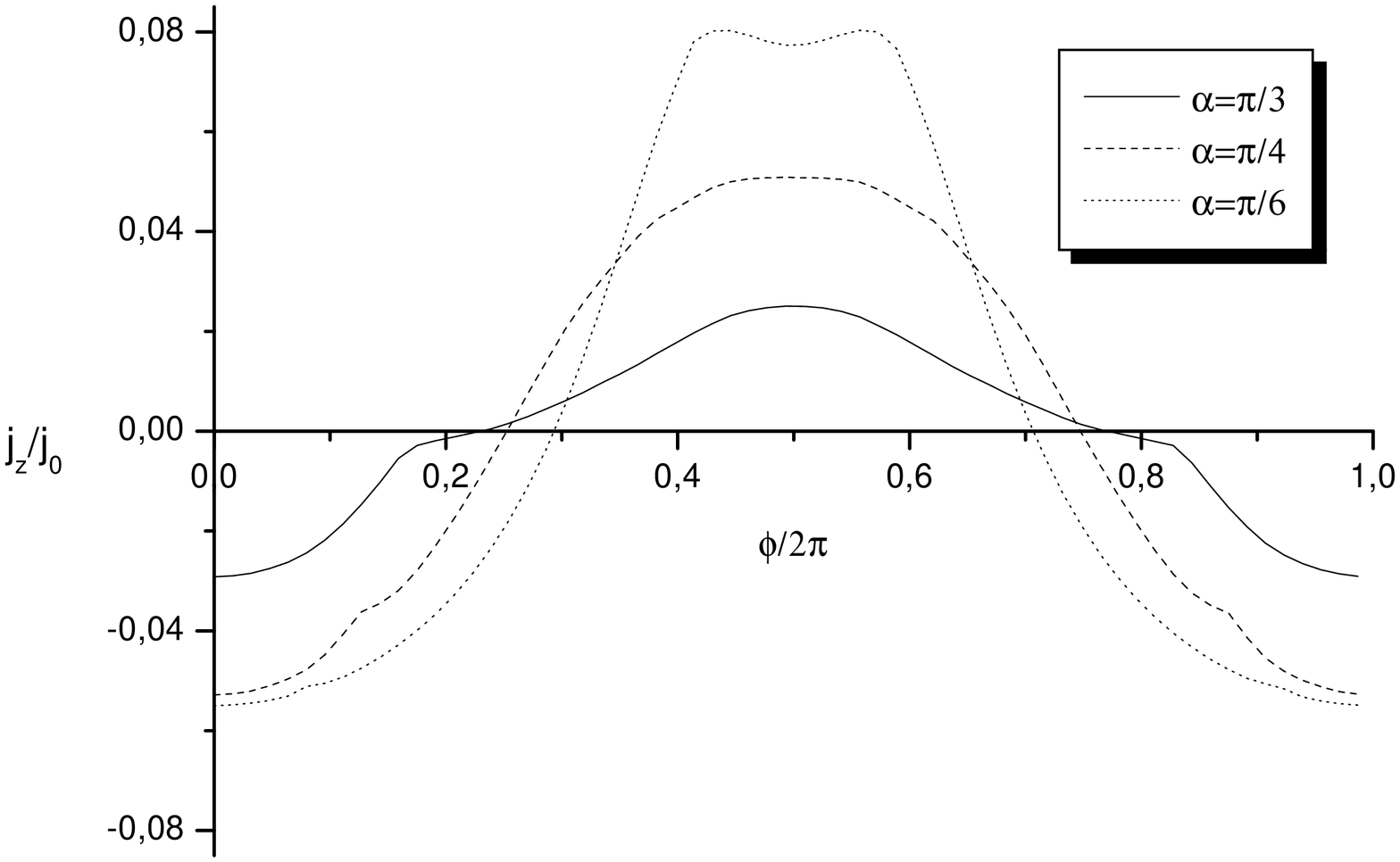}
\caption{Z-component of the tangential current versus phase
$\protect\phi $ for the axial state (22)in the geometry (ii) for
different $\protect\alpha $.}\label{Fig4}
\end{figure}

\begin{figure}[t]
\epsfysize 6cm \epsfbox[50 0 50 490]{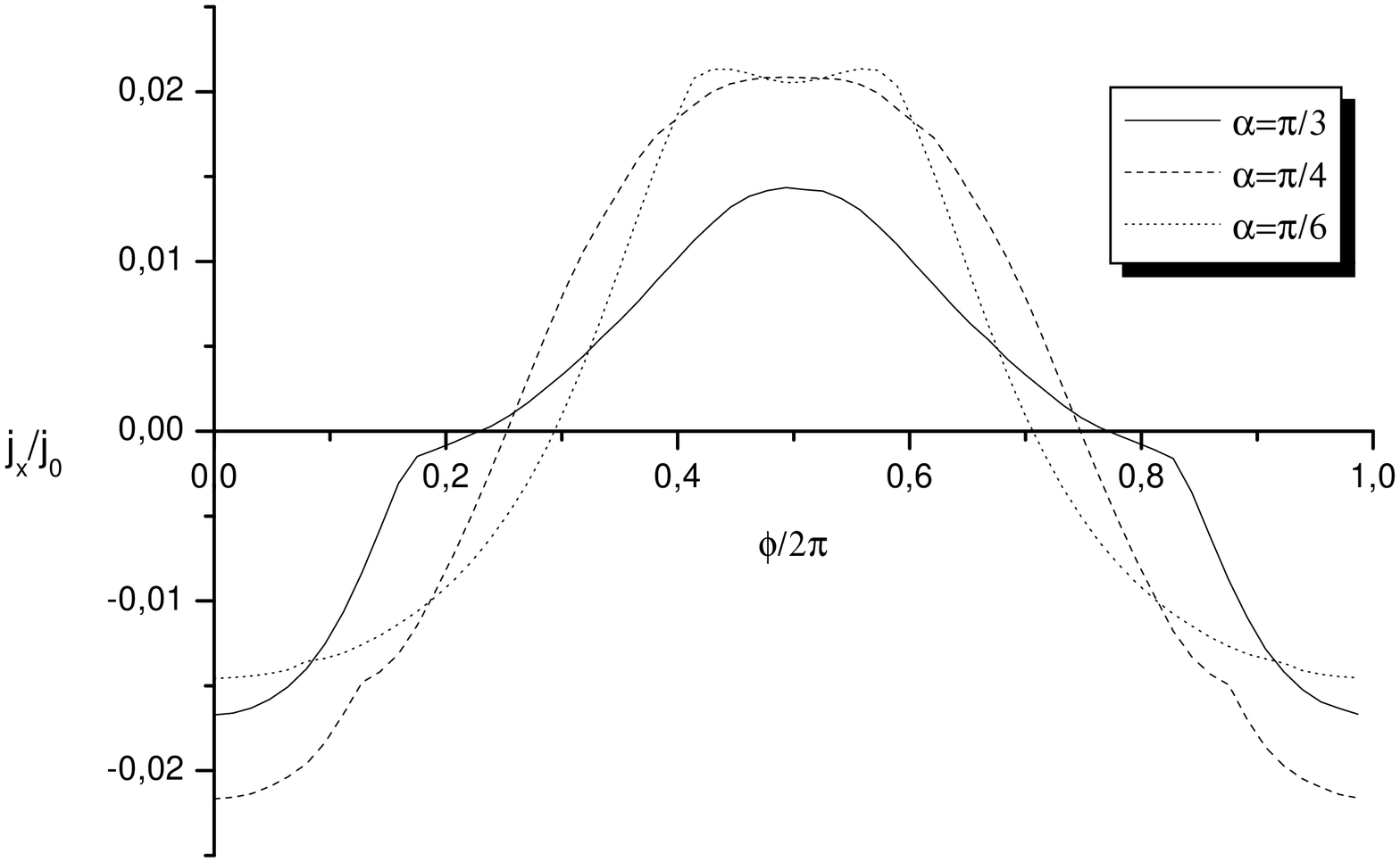}
\caption{X-component of the tangential current versus phase
$\protect\phi $ for the axial state (22)in the geometry (ii) for
different $\protect\alpha $.}\label{Fig5}
\end{figure}

Calculated $x$ and $z$-components of the current, which are
parallel to the surface, ${\bf j}_{S}(\phi )$ are shown in
Fig.\ref{Fig4} and Fig.\ref{Fig5} for the same axial state in the
geometry (ii). Note that the tangential to the surface current as
a function of $\phi$ is not zero when the Josephson current
(Fig.\ref{Fig3}) is zero. This spontaneous tangential current
(see also in Ref.\cite{BBF01}) is due to the specific ''proximity
effect'' similar to spontaneous current in contacts between
''$d$-wave'' superconductors \cite{AmOmZ,FY98_and_LSW00}. The
total current is determined by the Green function, which depends
on the order parameters in both superconductors. As a result of
this, for nonzero misorientation angles the parallel to the
surface current can be generated. In the geometry (i) the
tangential current for both the axial and planar states at $T=0$\
is absent.

The first candidate for the superconducting state in $Sr_{2}RuO_{4}$ was ''$%
p $-wave'' model \cite{Rice}\begin{equation} {\bf d=}\Delta
_{0}\widehat{z}\left( \widehat{k}_{x}+i\widehat{k}_{y}\right) .
\label{p-wave}
\end{equation}
Recently  \cite{Dahm,Graf1} it was shown that the
pairing state in $Sr_{2}RuO_{4},$ most likely, has lines of nodes.
It was suggested that this can occur if the spin-triplet state
belongs to a non trivial realization of the $E_{u}$
representation of $D_{4h}$ group, either $B_{2g}\otimes E_{u}$
\cite{Graf1} or $B_{1g}\otimes E_{u}$ \cite{Dahm} symmetry:
\begin{equation}
{\bf d=}\Delta _{0}\widehat{z}\widehat{k}_{x}\widehat{k}_{y}\left( \widehat{k%
}_{x}+i\widehat{k}_{y}\right) ,\text{ \ \ for }B_{2g}\otimes E_{u}\text{
symmetry;}  \label{hybrid2}
\end{equation}

\begin{equation}
{\bf d=}\Delta _{0}\widehat{z}\left( \widehat{k}_{x}^{2}-\widehat{k}%
_{y}^{2}\right) \left( \widehat{k}_{x}+i\widehat{k}_{y}\right) ,\text{ \ \
for }B_{1g}\otimes E_{u}\text{ symmetry.}  \label{hybrid1}
\end{equation}
Note that models (\ref{p-wave}-\ref{hybrid1})of the order
parameter spontaneously break time-reversal symmetry.

\begin{figure}[t]
\epsfysize 6cm \epsfbox[50 0 0 520]{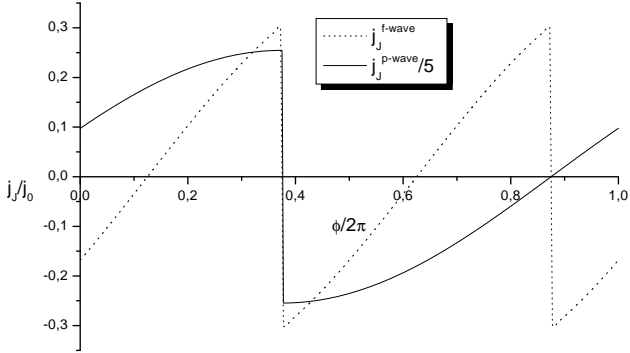}
\caption{\ {}Josephson current versus phase $\protect%
\phi $ for hybrid ''$f$-wave'' and ''$p$-wave'' states in the geometry (i); $%
\protect\alpha =\protect\pi /4$.} \label{Fig6}
\end{figure}

Taking into account a quasi-two-dimensional electron energy spectrum in $%
Sr_{2}RuO_{4}$, we calculate the current (\ref{jJos}) numerically
using a model of the cylindrical Fermi surface. The Josephson
current for the hybrid ''$f$-wave'' model of the order parameter
Eq.(\ref{hybrid1}) is compared
to  the $p$-wave model (Eq.\ref{p-wave}) in Fig.\ref{Fig6} (for $%
\alpha =\pi /4$). Note that the critical current for the
''$f$-wave'' model is several times smaller (for the same value of
$\Delta _{0}$) than for the ''$p$-wave'' model. This different
character of the current-phase dependencies enables us to
distinguish between the two states.

\begin{figure}[t]
\epsfysize 6cm \epsfbox[50 0 50 490]{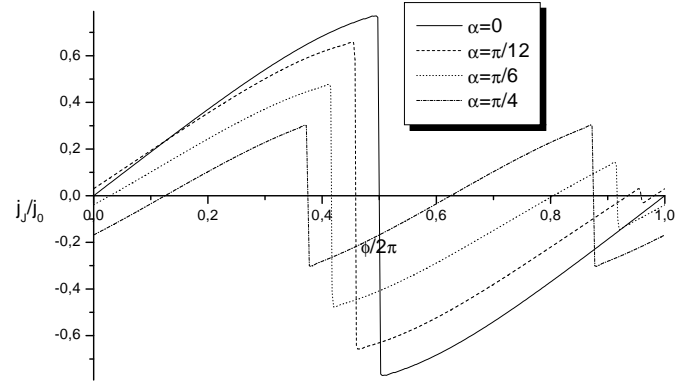} \caption{Josephson
current versus phase $\protect\phi $ for the
hybrid ''$f$-wave'' state in the geometry (i) for different $\protect%
\alpha $.} \label{Fig7}
\end{figure}

\begin{figure}[t]
\epsfysize 6cm \epsfbox[0 50 100 600]{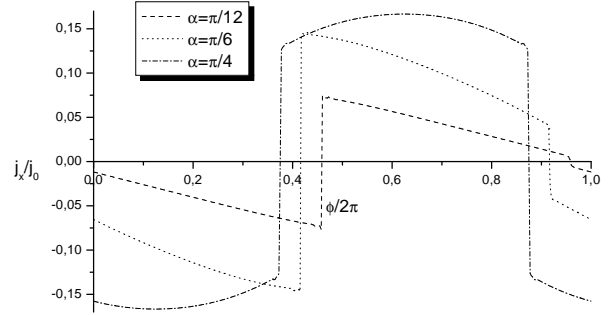}\caption{Tangential
current density versus phase $\protect\phi $ for the hybrid
''$f$-wave'' state in the geometry (i) for different
$\protect\alpha $.} \label{Fig8}
\end{figure}

In Figs.  \ref{Fig7} and \ref{Fig8} we present the Josephson
current and the tangential current for the hybrid ''$f$-wave''
model for different misorientation angles $\alpha $ (for the
''$p$-wave'' model it equals to the zero). Just as in Fig.2 for
the "$f$-wave" order parameter (22), in Fig.7 for the hybrid
''$f$-wave'' model (25) the steady state of the junction with
zero Josephson current corresponds to the nonzero spontaneous
phase difference if misorientation angle $\alpha\neq0$.


\section{Conclusion.}

We have considered the stationary Josephson effect in point-contacts between
triplet superconductors. Our consideration is based on models with ''$f$%
-wave'' symmetry of the order parameter belonging to the
two-dimensional representations of the crystallographic symmetry
groups. It is shown that the current-phase dependences are quite
different for different models of the order parameter. Because
the order parameter phase depends on the momentum direction on
the Fermi surface, the misorientation of the superconductors
leads to  spontaneous phase difference that corresponds to the
zero Josephson current and to the minimum of the weak link
energy. This phase difference depends on the misorientation angle
and can possess any values. We have found that in  contrast to
''$p$-wave'' model, in the ''$f$-wave'' models  the spontaneous
current may be generated in a direction which is tangential to
the orifice plane. Generally speaking this current is not equal
to zero in the absence of the Josephson current. We demonstrate
that the study of a current-phase dependence of small Josephson
junction for different crystallographic orientations of banks
enables one to judge on the
applicability of different models to the triplet superconductors $UPt_{3}$ and $%
Sr_{2}RuO_{4}.$

It is clear that such experiments require very clean
superconductors and perfect structures of the junction because of
pairbreaking effects of non-magnetic impurities and defects in
triplet superconductors. The influence of single impurities and
roughness of interface in the plane of the contact, which may
essentially decrease the critical current of the junction, will
be analyzed in our next paper.

\section{Acknowledgment.}

We would like to thank A.N. Omelyanchouk for many helpful
discussions. One of the authors (Yu.K) acknowledges to Institute
for Advanced Studies in Basic Sciences and personally Y.Sobouti
and M.R.H.Khajehpour for hospitality.

\end{document}